 \documentclass[final,5p,times]{elsarticle}

 \usepackage{graphicx}
\usepackage[normalem]{ulem}
\usepackage{amssymb}
\usepackage{wasysym}

\journal{Physics Letters B}

\begin{document}

\begin{frontmatter}


\title{Hypernuclear spectroscopy with K$^-$ at rest on $^7$Li, $^9$Be, $^{13}$C and $^{16}$O}


\author[polito,infnto]{M. Agnello}
\author[lnf]{L. Benussi}
\author[lnf]{M. Bertani}
\author[seul]{H. C. Bhang}
\author[unibs,infnpv]{G. Bonomi\corref{cor}}
\author[infnto,unito]{E. Botta}
\author[units,infnts]{M. Bregant\fnref{n1}}
\author[infnto,unito]{T. Bressani}
\author[infnto,unito]{S. Bufalino}
\author[infnto,unito2]{L. Busso}
\author[infnto]{D. Calvo}
\author[units,infnts]{P. Camerini}
\author[uniba,infnba]{B. Dalena}
\author[infnto,unito]{F. De Mori}
\author[uniba,infnba]{G. D'Erasmo}
\author[lnf]{F. L. Fabbri}
\author[infnto]{A. Feliciello}
\author[infnto]{A. Filippi}
\author[uniba,infnba]{E. M. Fiore}
\author[infnpv]{A. Fontana}
\author[kyoto]{H. Fujioka}
\author[unipv]{P. Genova}
\author[lnf]{P. Gianotti}
\author[infnts]{N. Grion}
\author[lnf]{V. Lucherini}
\author[infnto,unito]{S. Marcello}
\author[teheran]{N. Mirfakhrai}
\author[unibs,infnpv]{F. Moia}
\author[infnpv,unipv]{P. Montagna}
\author[inaf,infnto]{O. Morra}
\author[kyoto]{T. Nagae}
\author[riken]{H. Outa}
\author[infnba]{A. Pantaleo\fnref{n2}}
\author[infnba]{V. Paticchio}
\author[infnts]{S. Piano}
\author[units,infnts]{R. Rui}
\author[uniba,infnba]{G. Simonetti}
\author[infnto]{R. Wheadon}
\author[unibs,infnpv]{A. Zenoni}

\cortext[cor]{ Corresponding author: G. Bonomi (germano.bonomi@ing.unibs.it).}
\fntext[n1]{ Now at SUBATECH, Ecole des Mines de Nantes, Universit\'e de Nantes, CNRS-IN2P3, Nantes, France.}
\fntext[n2]{ Deceased. We wish to dedicate the present Letter to his memory.}

\address[polito] {Dipartimento di Fisica, Politecnico di Torino, Corso Duca degli Abruzzi 24, Torino, Italy}
\address[infnto] {INFN Sezione di Torino, via P. Giuria 1, Torino, Italy }
\address[lnf] {Laboratori Nazionali di Frascati dell'INFN, via. E. Fermi, 40, Frascati, Italy }
\address[seul] {Department of Physics, Seoul National University, 151-742 Seoul, South Korea }
\address[unibs] {Dipartimento di Ingegneria Meccanica e Industriale, Universit\`a di Brescia, via Branze 38, Brescia, Italy }
\address[infnpv] {INFN Sezione di Pavia, via Bassi 6, Pavia, Italy }
\address[unito] {Dipartimento di Fisica Sperimentale, Universit\`a di Torino, Via P. Giuria 1, Torino, Italy }
\address[units] {Dipartimento di Fisica, Universit\`a di Trieste, via Valerio 2, Trieste, Italy }
\address[infnts] {INFN Sezione di Trieste, via Valerio 2, Trieste, Italy }
\address[unito2] {Dipartimento di Fisica Generale, Universit\`a di Torino, Via P. Giuria 1, Torino, Italy }
\address[uniba] {Dipartimento di Fisica Universit\`a di Bari, via Amendola 173, Bari, Italy }
\address[infnba] {INFN Sezione di Bari, via Amendola 173, Bari, Italy }
\address[kyoto] {Department of Physics, Kyoto University, Sakyo-ku, Kyoto Japan }
\address[unipv] {Dipartimento di Fisica Nucleare e Teorica, Universit\`a di Pavia, via Bassi 6, Pavia, Italy }
\address[teheran] {Department of Physics, Shahid Behesty University, 19834 Teheran, Iran }
\address[inaf] {INAF-IFSI, Sezione di Torino, corso Fiume 4, Torino, Italy }
\address[riken] {RIKEN, Wako, Saitama 351-0198, Japan }

\begin{abstract}
The FINUDA experiment collected data to study the production of hypernuclei on different nuclear targets. The hypernucleus formation occurred through the strangeness-exchange reaction $K^-_{stop} + \; ^AZ \rightarrow \; ^A_{\Lambda}Z + \pi^-$. From the analysis of the momentum of the emerging $\pi^-$, binding energies and formation probabilities of  $^7_{\Lambda}$Li, $^9_{\Lambda}$Be, $^{13}_{\Lambda}$C and $^{16}_{\Lambda}$O have been measured and are here presented. The behavior of the formation probability as a function of the atomic mass number A is also discussed.

\end{abstract}


\end{frontmatter}






\bibliographystyle{elsarticle-num}
\bibliography{<your-bib-database>}

\begin{thebibliography}{00}
%
\bibitem{uno}
M. Danysz, J. Pniewski, Philos. Mag. {\bf 44} (1953) 348.
%
\bibitem{hashimoto}
O. Hashimoto and H. Tamura, {\it Spectroscopy of $\Lambda$ hypernuclei}, Prog. Part. and Nucl. Phys. {\bf 57} (2006) 564-653.
%
\bibitem{cusanno}
F. Cusanno {\it et al.}, Phys. Rev. Lett. {\bf 103} (2009) 202501.
%
\bibitem{iodice}
M. Iodice {\it et al.}, Phys. Rev. Lett. {\bf 99} (2007) 052501.
%
\bibitem{bressani}
T. Bressani, in Proceedings of the Workshop on Physics and Detectors for DA$\Phi$NE, Frascati, 1991, edited by G. Pancheri (Laboratori Nazionali di Frascati, Frascati, 1991), p. 475.
%
\bibitem{faessler}
A. M. Faessler {\it et al.}, Phys. Lett. B {\bf 46} (1973) 468.
%
\bibitem{hayano}
H. Tamura, R. S. Hayano, H. Outa and T. Yamazaki, Prog. Theor. Phys. Suppl. {\bf 117} (1994) 1.
%
\bibitem{ahmed}
M. W. Ahmed {\it et al.}, Phys. Rev. C {\bf 68} (2003) 064004
%
\bibitem{C12}
M. Agnello {\it et al.}, Phys. Lett. B {\bf 622} (2005) 35.
%
\bibitem{MWD}
M. Agnello {\it et al.}, Phys. Lett. B {\bf 681} (2009) 139.
%
\bibitem{NMWD}
M. Agnello {\it et al.}, Phys. Lett. B {\bf 685} (2010) 247.
%
\bibitem{teo1}
J. H\"ufner, S.Y. Lee, and H.A. Weidenm\"uller, Nucl. Phys. A {\bf 234} (1974) 429. 
%
\bibitem{teo2} 
H. R. Dalitz and A. Gal, Ann. of Phys. {\bf 116} (1978) 167.
\bibitem{teo2bis}
A. Gal and L. Klieb, Phys. Rev. C {\bf 34} (1986) 956. 
%
\bibitem{teo2ter}
D.J. Millener {\it et al.}, Phys. Rev. C {\bf 31} (1985) 499, \\
D.J. Millener, Nucl. Phys. A {\bf 450} (1986) 199c.
\bibitem{teo3}
A. Matsuyama and K. Yazaki, Nucl. Phys. A {\bf 477}, 673 (1988). 
%
\bibitem{teo4}
 A. Ciepl«y, E. Friedman, A. Gal, and J. Mareÿs, Nucl. Phys. A {\bf 696}, 173 (2001). 
%
\bibitem{teo5}
V. Krej$\mathrm{\check{c}i\check{r}}$\'ik, A. Ciepl\'y and A. Gal, Phys. Rev. C {\bf 82} (2010) 024609. 
%
\bibitem{AME}
A. H. Wapstra, G. Audi, and C. Thibault Nucl. Phys. A {\bf 729} (2003), 129;\\ 
G. Audi, A. H. Wapstra, and C. Thibault. Nucl. Phys. A {\bf 729} (2003), 337.
\bibitem{abramov}
B. M. Abramov {\it et al.}, JETP Letters {\bf 71} (2000) 359.
%
\bibitem{minuit}
F. James and M. Roos, Computer program MINUIT, CERN program library (1989) writeup CERN D506.
%
\bibitem{TFF}
R. Barlow and C. Beeston, Comp. Phys. Comm. {\bf 77} (1993) 219-228.
%
\bibitem{root}
Rene Brun and Fons Rademakers, Nucl. Inst. and Meth. in Phys. Res. A {\bf 389} (1997) 81-86 (see also http://root.cern.ch).
%
\bibitem{pdg}
K. Nakamura {\it et al.} (Particle Data Group), J. Phys. G {\bf 37}, (2010) 075021.
%
\bibitem{li7}
M. Agnello {\it et al.}, Proceedings of the IX International Conference on Hypernuclear and Strange Particle Physics 
HYP 2006 October, Mainz, Germany, 57.
%
\bibitem{davis}
D. H. Davis {\it et al.}, Nucl. Phys. A {\bf 754} (2005) 3c. 
%
\bibitem{ukaili7}
M. Ukai et al., Phys. Rev. C {\bf 73} (2006) 012501(R).
%
\bibitem{tam}
H. Tamura {\it et al.}, Nucl. Phys. A  {\bf 754} (2005) 58c.
%
\bibitem{itonaga}
K. Itonaga, T. Motoba and H. Bando, Prog. Theor. Phys.  {\bf 84} (1990) 291.
%
\bibitem{koh}
H. Kohri {\it et al.}, Phys. Rev. C  {\bf 65} (2002) 034607.
%
\bibitem{millener}
D. J. Millener, Nucl. Phys. A {\bf 804} (2008) 84.
%
\bibitem{uka}
M. Ukai {\it et al.}, Phys. Rev. C  {\bf 77} (2008) 054315.
%
\bibitem{cieply}
A. Ciepl\'y, E. Friedman, A. Gal and V. Krej$\mathrm{\check{c}i\check{r}}$\'ik, following Letter. 
%

\end{thebibliography}


\section{Introduction}
\label{intro}
FINUDA stands for $FI$sica $NU$cleare a $DA$$\Phi$NE, that is Nuclear Physics at DA$\Phi$NE, the $e^+$-$e^-$ collider of the INFN ``Laboratori Nazionali di Frascati'', close to Rome. One of the main aims of the experiment was the study of production and decay of $\Lambda$-hypernuclei. The creation of a hypernucleus \cite{uno}, that is a nucleus in which a nucleon is replaced by a hyperon (for example a neutron is substituted by a $\Lambda$), requires the injection of ${\it strangeness}$ into the nucleus. This is possible in different ways (see \cite{hashimoto} and references therein for details), mainly using a $\pi^+$ or a $K^-$ beam on fixed targets; recently also electron beams have been used \cite{cusanno, iodice}.
The use of meson beams ($\pi^+$ and $K^-$) usually requires relatively  thick targets (few g/cm$^2$) in order to allow the particle a reasonable interaction rate, reducing thus the instrumental resolution. The original idea of FINUDA \cite{bressani} was to use a particular feature of the DA$\Phi$NE machine, where
the $e^+$-$e^-$ beams circulate with an energy of 510 MeV in order to produce the $\Phi(1020)$ meson in the head-on collisions. This particle 
decays, with a branching ratio of 49.2\%, in two back-to-back kaons ($\Phi \rightarrow K^+K^-$) with kinetic energy as low as $\sim$16 MeV. In this way an unconventional  monochromatic source of very low energy $K^-$ was available for the experiment, allowing the use of much thinner targets, 0.1-0.2 g/cm$^2$ compared to some g/cm$^2$ of previous hypernuclear fixed-target experiments. In FINUDA the $K^-$'s were slowed down to rest in the targets leading to the production of $\Lambda$ hypernuclei through the strangeness-exchange reaction:
\begin{equation}
K^-_{stop} + ^AZ \rightarrow ^A_{\Lambda}Z + \pi^- 
\label{e:prod}
\end{equation}
where $^AZ$ indicates the target nucleus and $^A_{\Lambda}Z$ the $\Lambda$ hypernucleus in which a $\Lambda$ particle replaced a neutron. By precisely measuring the momentum of the outgoing pions, it is possible to determine the energy levels of the produced hypernucleus, and by counting them the hypernucleus formation probability can be deduced. Up to now only few measurements of formation probability have been performed. Following the first experiment on a $^{12}$C stopping target \cite{faessler}, measurement on some other nuclei were subsequently performed \cite{hayano}. A low statistics measurement on the ($K^-_{stop}, \pi^o$) reaction on $^{12}$C was recently published \cite{ahmed}. A first measurement by FINUDA on a $^{12}$C target was also previously reported \cite{C12}. Other information about hypernuclear state formation can be found in recent FINUDA publications on mesonic and non mesonic hypernuclei decay \cite{MWD, NMWD}. 
Theoretical calculations \cite{teo1,teo2,teo2bis,teo2ter,teo3,teo4} have mostly reported formation rates which are substantially lower than those reported in previous measurements, and this holds true also for the recent \cite{teo5} when compared with the formation rates reported in the present work. In the following, after a brief description of the FINUDA experimental apparatus, details on the data analysis and the results on hypernucleus formation probabilities and binding energies for $^7_\Lambda$Li, $^9_\Lambda$Be, $^{13}_\Lambda$C and $^{16}_\Lambda$O are presented. 

\section{The FINUDA experiment}
Differently from previous hypernuclear fixed target experiments, FINUDA had an unconventional geometry, typical of collider experiments. 
The whole apparatus was contained inside a superconducting solenoid which provided a homogeneous magnetic field of 1.0 T over a cylindrical volume of 146 cm in radius and 211 cm in length and it had been designed to obtain a large acceptance of about 2$\pi$ sr around the beam interaction region. More detailed descriptions of the detector may be found in \cite{C12, MWD, NMWD} and references therein. The particles coming from the ($e^+$-$e^-$) interaction point travelled radially outwards encountering three main apparatus regions ({\it target}, {\it tracking} and {\it time of flight}). In the first one, the {\it target region}, a barrel of 12 thin scintillators, called TOFINO, surrounded the beam pipe and detected the back-to-back $K^- - K^+$ coming from the $\Phi$ decay. The signals from these detectors were used for triggering purposes. The scintillators were surrounded by an octagonal array of double-sided silicon microstrip detectors, called ISIM and having a spatial resolution better than 30 $\mu$m and a good energy resolution ($\Delta$E/$E$ of 20\% for the low energy kaons from the $\Phi$ decay) within a wide dynamic range (up to $\sim$ 20 MIP's). They traced the kaons before entering the 8 target modules facing the silicon detectors at a distance of a couple of millimeters. The apparatus could study simultaneously up to 8 different targets, thus reducing the difficulties in comparing results from different elements. The kaon stopping point was determined with a resolution of about 800 $\mu$m due to the angular and energy straggling. The charged particles emitted after kaon absorption were traced in the {\it tracking region}. First they were detected by another array of ten silicon microstrip detectors (OSIM) placed close to the targets. OSIM was also used for particle identification purposes thanks to its good energy resolution. Moving  radially outwards two arrays of eight planar low-mass drift chambers (LMDC) provided the measurement of the particle trajectories with a spatial resolution of $\sigma_{\rho\Phi} \sim 150 \;\mu$m and $\sigma_{z} \sim 1$ cm. The outer tracking device consisted of six layers of longitudinal and stereo straw tubes with a spatial resolution of $\sigma_{\rho\Phi} \sim 150 \;\mu$m and $\sigma_{z} \sim 500 \;\mu$m. The latter region ({\it time of flight}) enclosed the FINUDA experimental apparatus. 
It was composed of a barrel, called TOFONE, of 72 scintillator slabs (10 cm wide and 255 cm long) that provided signals for the first level trigger, for the time-of-flight measurement of the charged particles and for the detection of neutrons.\\

FINUDA collected data in two different periods. The data discussed in the following were accumulated in the second data taking lasted from November 2006 to June 2007 with a maximum daily luminosity of about 10 pb$^{-1}$ and a total of 966 pb$^{-1}$. The targets were two of $^6$Li, two of $^7$Li, two of $^9$Be, one of $^{13}$C and one of D$_2$O. The trigger selected events with two signals above the kaon detection threshold in two back-to-back TOFINO slabs within a time coincidence with a TOFONE barrel signal. 

\section{Data analysis}
\begin{figure}[t]
\begin{center}
\resizebox{7.5cm}{!}{\includegraphics{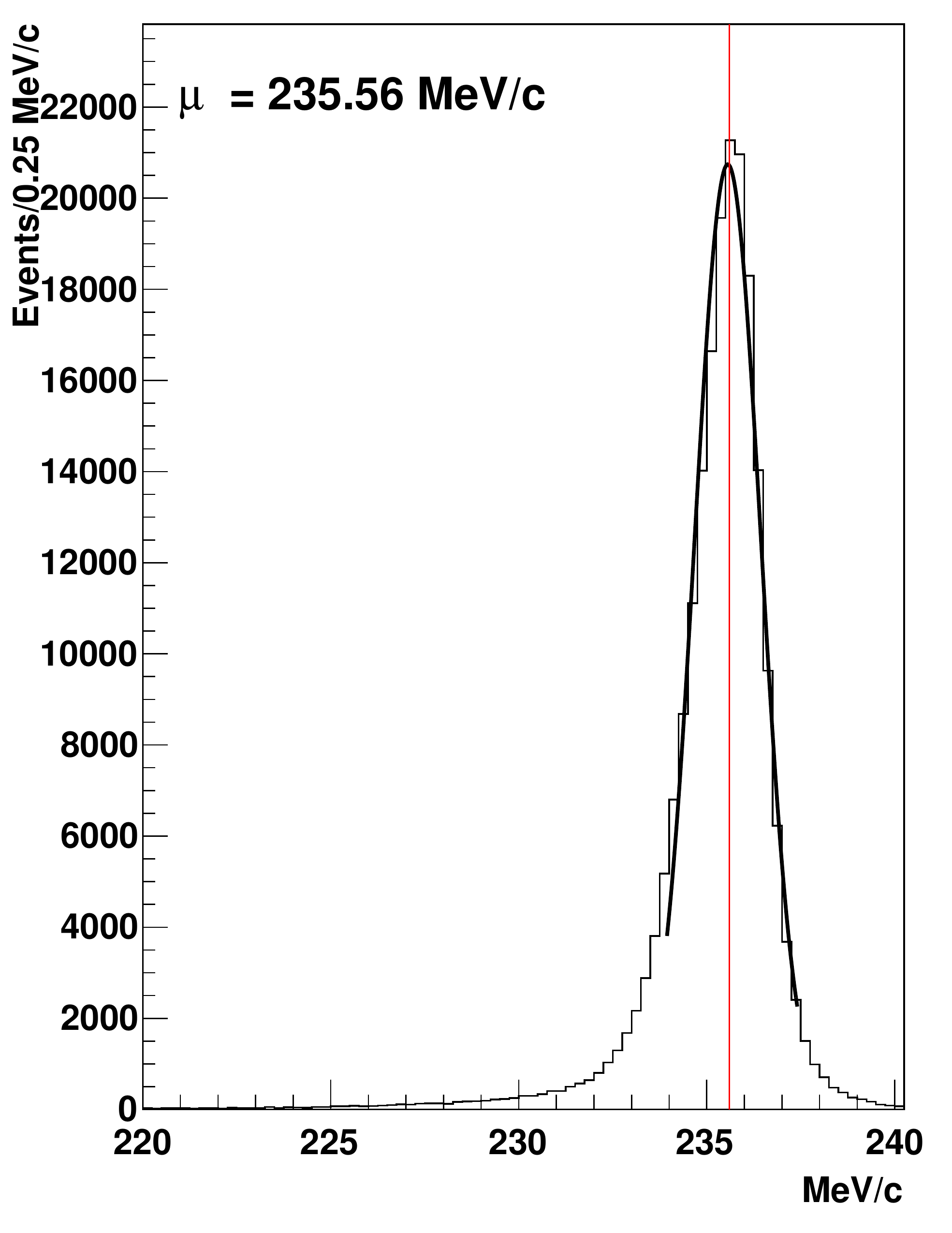}}
\caption{Momentum distribution of muons coming from the decay of the $K^+$'s from a single target. The expected value of 235.6 MeV/c is represented with a line.}
\label{f:mu}
\end{center}
\end{figure}
\begin{figure}[t]
\begin{center}
\resizebox{7.5cm}{!}{\includegraphics{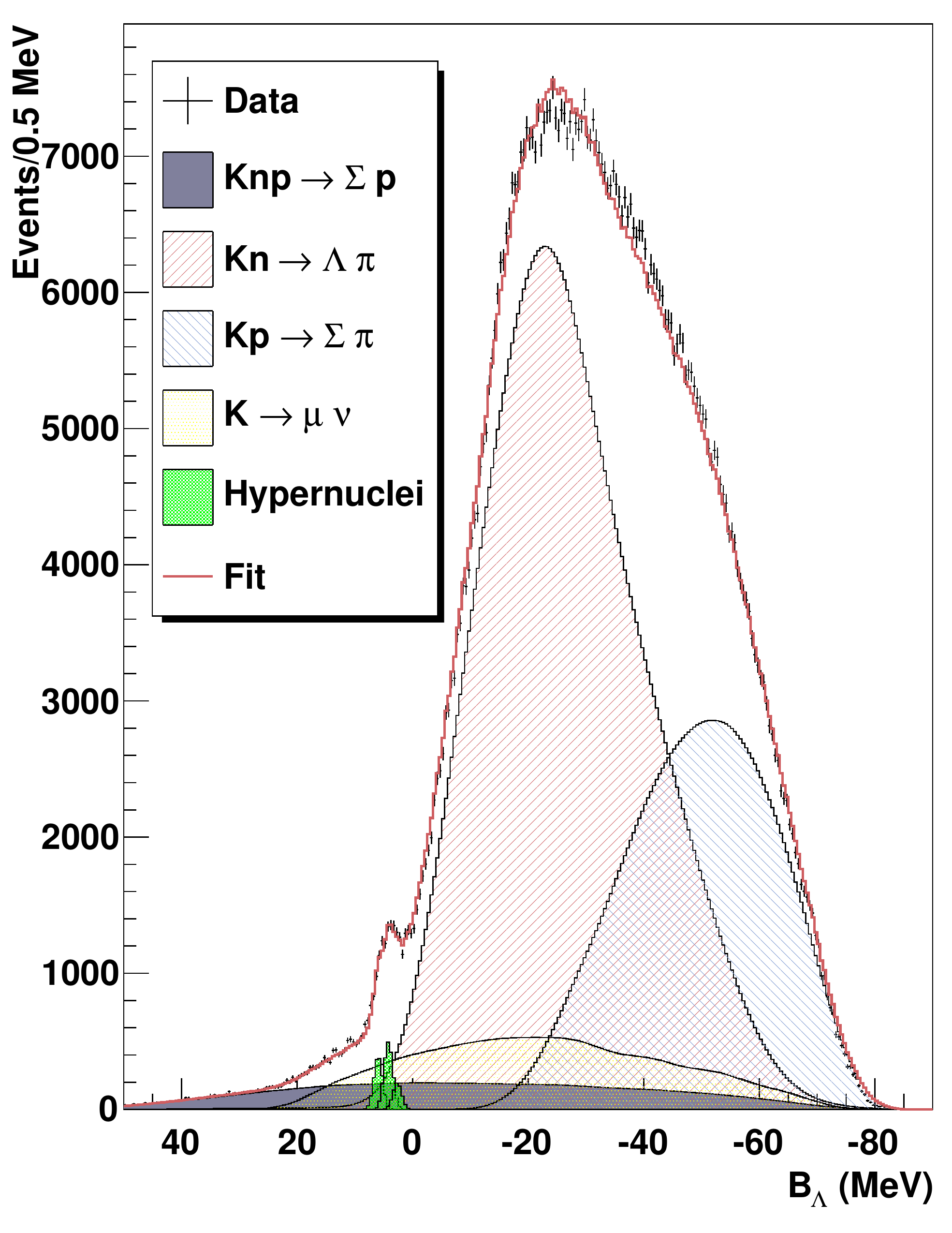}}
\caption{Binding energy distribution for the $^7$Li targets. The superimposed fit is the sum of the four background distributions and of three Gaussians for the hypernuclear signal (see text for details).}
\label{f:li7all}
\end{center}
\end{figure}

The kaons coming from the decay of the $\Phi$ meson, when flying outwards, are slowed down in the beam pipe, in the TOFINO scintillators and in the ISIM silicon detectors before encountering one of the 8 targets. The target thickness, of the order of some mm depending on the material density (2 mm for $^9$Be, 3 mm for D$_2$O, 4 mm for $^7$Li and 1 cm for $^{13}$C), 
has been chosen so as to stop the kaons inside the target itself close to the external surface. The stopping point of both $K^+$ and $K^-$ is calculated simultaneously using the information of the ISIM silicon detector and imposing a double helix trajectory with a common origin in the $\Phi$ decay vertex. Since the $K^+$ decays more abundantly into a $\mu^+ + \nu_{\mu}$ with a well known branching ratio, this reaction provides in FINUDA a unique reference both for energy and absolute rate measurements as will be discussed with more details later. Once stopped the $K^-$ can undergo a number of reactions among which the above mentioned strangeness-exchange reaction (\ref{e:prod}). The data analysis performed in the present work is based on the reconstruction of the $\pi^-$ tracks emerging from the $K^-$ stopping points. Tracks with a signal in OSIM, in both layers of the drift chambers and in the straw tubes, and satisfying quality cuts such as a good $\chi^2$ and a minimum extrapolated distance from the kaon stopping point have been selected and used. In the following we may refer to {\it forward} and {\it backward} tracks depending on how the track exits the target. If it flies outwards it is called forward, otherwise, when it crosses back the beam pipe region, it is called backward. Another difference is the distinction between pions generated from targets in the  {\it boost-side} and from the {\it antiboost-side}. Indeed the DA$\Phi$NE $e^+$-$e^-$ interaction is not completely collinear, the two beams colliding with an angle of 12.5 mrad. This causes the kaons exiting in one hemisphere to have slightly more energy with respect to those going in the opposite hemisphere. This distinction will be useful when discussing the background modeling for the signal extraction.
Once the $\pi^-$ track has been reconstructed, the pion momentum can be measured with a resolution $\Delta p/p$ that varies from 0.5 $\%$ to 0.9 $\%$ depending on the vertex position and on the quality cuts required. The resolution can be evaluated measuring the width of the momentum distribution of the $\mu^+$'s coming from the decay of the $K^+$ (fig.\ref{f:mu}). \\
From the value of the $\pi^-$ momentum and imposing the energy and the momentum conservation laws, the hypernucleus mass in a specific level can then be calculated as:
\begin{equation}
m_{Hyp,i} = \sqrt{(m_{K^-} + m_{^A{\rm Z}} - E_{\pi,i})^2 - p_{\pi,i}^2} 
\label{e:mhyp}
\end{equation}
where $m_{K^-}$ is the $K^-$ mass, $m_{^A{\rm Z}}$ is the target nucleus mass in the ground state, $m_{Hyp,i}$ the mass of the particular  $^A_{\Lambda}$Z hypernucleus formed in the i$^{th}$ energy level state, $p_{\pi,i}$ is the pion momentum for the produced hypernucleus level and $E_{\pi,i}$ the corresponding pion total energy. In general, in the literature, the $\Lambda$ binding energy $B_{\Lambda,i}$ is often used. It is defined by the relationship 
\begin{equation}
 B_{\Lambda,i}  = (m_{^{A-1}{\rm Z}} + m_{\Lambda}) - m_{Hyp,i} 
\label{e:bind}
\end{equation}
where $m_{^{A-1}{\rm Z}}$ indicates the mass of the hypernuclear core in its ground state and $m_{\Lambda}$ the mass of the $\Lambda$ particle. For the calculations of the nuclear masses the AME2003 table has been used \cite{AME}.  
An example of $B_{\Lambda}$ distribution is shown in fig.~\ref{f:li7all} for the $^7$Li targets. The bump in the bound region corresponds to the production of hypernuclei. As it can be clearly seen from fig.~\ref{f:li7all}, the overall behaviour of the experimental data, not corrected for acceptance, has been reproduced with the sum of contributions from background reactions, as explained in details in the following, and of Gaussian functions for the signals.
Indeed when a $K^-$ stops inside a given nucleus of the target, beyond producing a hypernucleus, it can undergo a number of possible nuclear reactions. 
The major contribution of the pion spectra coming from $K^-$ nuclear absorption is the production of $\Lambda$ and $\Sigma$ hyperons, by strangeness exchange reactions, and their decays. Only if the $\Lambda$ hyperon is captured by the nucleus, the hypernucleus can be formed. By means of a realistic MC simulation of Quasi Free $\Lambda$ and $\Sigma$ production and decay in FINUDA it was possible to determine the processes that contribute, among others, as a background in the $B_{\Lambda}$ spectra:\\
(I) $K^- (np) \rightarrow \Sigma^- p$  (followed by $\Sigma^- \rightarrow n \pi^-$ decay) \\
(II) $K^- n \rightarrow \Lambda \pi^- $ (so called {\it $\Lambda$-Quasi Free})\\
(III) $K^- p \rightarrow \Sigma^- \pi^+$ (followed by $\Sigma^- \rightarrow n \pi^-$ decay) \\
Concerning the Quasi Free $\Lambda$ production, whose shape is particularly sensitive to the formation probability of the hypernuclear states around the threshold, two models were used for 
the nucleons Fermi momentum distribution, the ``Fermi-gas'' model and the ``oscillator'' model \cite{abramov}. Specific simulations using distinct mixtures of the two models were done target 
by target to best reproduce the experimental spectra.
Another process contributing to the background for the {FINUDA} hypernuclear data was found to be \\
(IV) $K^- \rightarrow \mu^- \bar{\nu}_{\mu}$\\
that is the decay in flight of the $K^-$ occurring close to the target. The $\mu^-$ produced by such decay can be reconstructed as a $\pi^-$ and can give an entry in the signal region. This background proved to be significant for forward tracks emitted from the boost-side targets and negligible for backward tracks and antiboost-side targets.
All the four background reactions have been generated with the FINUDA Monte Carlo and reconstructed with the FINUDA reconstruction program in order to reproduce the background sitting below the hypernuclear signal. The experimental data have been thus fitted to the sum of the histograms representing the relevant backgrounds, and of Gaussians, whose number depends on the number of peaks seen in the experimental distributions, for the signal (see fig.~\ref{f:li7all}). 
The reliability of the background shapes used is supported by the fact that the binding energy distribution is well reproduced in its integrity. The weights of the background reactions have been left free to vary. The width of the Gaussians was the same for all the peaks and has been fixed at 
$\sigma = $ 0.75 MeV corresponding to 1.76 MeV FWHM. The fit has been performed in two steps; in the first one the MINUIT algorithm \cite{minuit} was used to fix the mean value of the Gaussians, while in the second one a more sophisticated algorithm \cite{TFF}, implemented in the TFractionFitter class of ROOT \cite{root}, was employed. \\

The position of the mean of the Gaussians in the $B_{\Lambda}$ spectra, the one at the highest value representing the ground state and the others the excited states, provides the binding energy of the hypernuclear states, while the number of reconstructed events in the peaks is related to the probability of formation of a given hypernuclear state. The measurement of such absolute rate implies an accurate determination of the apparatus acceptance as well as of trigger and reconstruction efficiencies for the considered reaction. The hypernuclear formation probability $R_{hyp}$ per $K^-$ stopped inside the target can be extracted from the following relation: $n_{hyp} = N_{K^-} \cdot R_{hyp} \cdot \epsilon_D \cdot \epsilon_{\pi}$ where $n_{hyp}$ is the number of hypernuclei detected (number of events in the Gaussians of the experimental data fit), $N_{K^-}$ the number of detected $K^-$ stopped inside the given target (of the order of 10 millions per target), $\epsilon_D$ the efficiency in detecting the $\pi^-$ track (correlated to the detector efficiencies) and $\epsilon_{\pi}$ the efficiency in reconstructing the $\pi^-$ (correlated to the trigger bias, reconstruction algorithm and selection cuts). The formation probability can thus be calculated with the following formula: 
\begin{equation}
R_{hyp}  = \frac{n_{hyp}}{N_{K^-}}  \cdot \frac{1}{\epsilon_D} \cdot \frac{1}{\epsilon_{\pi}}
\label{e:fp}
\end{equation}
While the number of pions $n_{hyp}$ and the number of stopped kaons $N_{K^-}$ come from counting the events with required features, the evaluation of $\epsilon_D$ and $\epsilon_{\pi}$ needs the help of Monte Carlo simulations and of other experimental data.\\
The efficiency $\epsilon_{\pi}$ has been calculated simulating the formation of hypernuclei, generating hypernuclear events with a fixed known probability $R_{hyp}^{MC}$ along with background events with $\pi^-$ momentum distribution similar to the experimental one. The data have been then reconstructed assuming detector efficiencies of 100 $\%$. In this way $\epsilon_{\pi}$ could be calculated using the relationship $\epsilon_{\pi} = \frac{n_{hyp}^{MC}}{N_{K^-}^{MC}}  \cdot \frac{1}{R_{hyp}^{MC}}$. A number of events similar to the experimental one has been generated.\\
For the calculation of  $\epsilon_D$ FINUDA could exploit the detected $\mu^+$ coming from the $K_{\mu 2}$ decay process ($K^+\rightarrow \mu^+ \nu_{\mu}$), whose branching ratio is well known to be BR($K_{\mu 2}$) = 63.55$\%$ \cite{pdg}. This decay provides a physical reference rate to which the unknown hypernuclear rates can be referred. The number of detected and reconstructed $\mu^+$ ($n_{\mu^+}$) is correlated to the number of stopped $K^+$ inside a target ($N_{K^+}$) by the relationship $n_{\mu^+} = N_{K^+} \cdot \mathrm{BR}(K_{\mu 2}) \cdot \epsilon_D \cdot \epsilon_{\mu}$, where $\epsilon_D$ is the detector efficiency and $\epsilon_{\mu}$, analogously to $\epsilon_{\pi}$, is the efficiency in reconstructing the $\mu^+$. The value of $\epsilon_D$ has been extracted from the experimental data counting how many muons we were able to reconstruct for each stopped $K^+$ and accounting for the decay branching ratio: $\epsilon_D = \frac{n_{\mu^+}}{N_{K^+} \cdot \mathrm{BR}(K_{\mu 2})} \cdot \frac{1}{\epsilon_{\mu}}$. Since the curvature of the positive muons and of the negative pions is the opposite due to their charge, the {\it path} followed by $\mu^+$ and $\pi^-$ when exiting a given target is different and it can intersect different set of detectors. This effect has been taken carefully into account. Overall detector efficiencies from 45 $\%$ to 65 $\%$, depending of the target position, have been calculated.\\
Given the above formula (\ref{e:fp}), for each target, the value of formation probability has been calculated for different sets of quality cuts, for different bin sizes (from 0.25 MeV to 1 MeV per bin) and also for the event sample in which also the $\mu^+$ coming from the $K^+$ decay was reconstructed. For this selection of events, called {\it $\mu$-tag}, the trigger could have been given with high probability by the muon itself instead of the negative pion. This helped to determine the systematic error in evaluating the trigger efficiency.\\

The formation probability reported in the following is referred to the average over all the different calculations (quality cuts, bin sizes, trigger selection). Each value is presented with two errors. The first one comes from a typical error propagation for the formula (\ref{e:fp}) and is a combination of statistical and systematic errors. The second one takes into account the difference between the average value of all the different calculations for the complete set of events and for the $\mu$-tag sub-sample. This systematic error accounts for the uncertainties in simulating the FINUDA trigger. While the first error changes from target to target, depending upon the statistics, the detector efficiencies 
and the background shapes, the second one, in percentage, is the same for all the targets and amounts to 14 $\%$. This error is clearly needed when comparing a single target result with other experiments or theoretical calculations, but should be ignored when using the formation probabilities here reported to evaluate the A dependence or when using the ratio between different targets. Indeed this error would move all the values up or down by the same amount. \\

For what concerns the error in determining the position of the Gaussian peaks, thus in measuring the hypernuclei binding energy, once again FINUDA has the big advantage of having a self-calibrating system. The muons coming from the $K^+$ decay are centered around the peak value of 235.6 MeV/c. In fig.~\ref{f:mu} the momentum distribution of muons coming from a single target is shown. The difference between the expected value and the measured one varies from target to target with an upper limit at about 0.3 MeV/c. We can thus state that the absolute energy scale is known with a precision of $\sim$ 0.3 MeV. In addition when performing the fit of the hypernuclear peaks for the different sub-samples (varying as said before quality cuts, bins, trigger requirement, etc.), we also saw variations up to 0.3 MeV. Summing quadratically these two types of uncertainties an overall error of 0.4 MeV has been calculated to affect our measurements of binding energies. 

\begin{figure}[t]
\begin{center}
\resizebox{7.5cm}{!}{\includegraphics{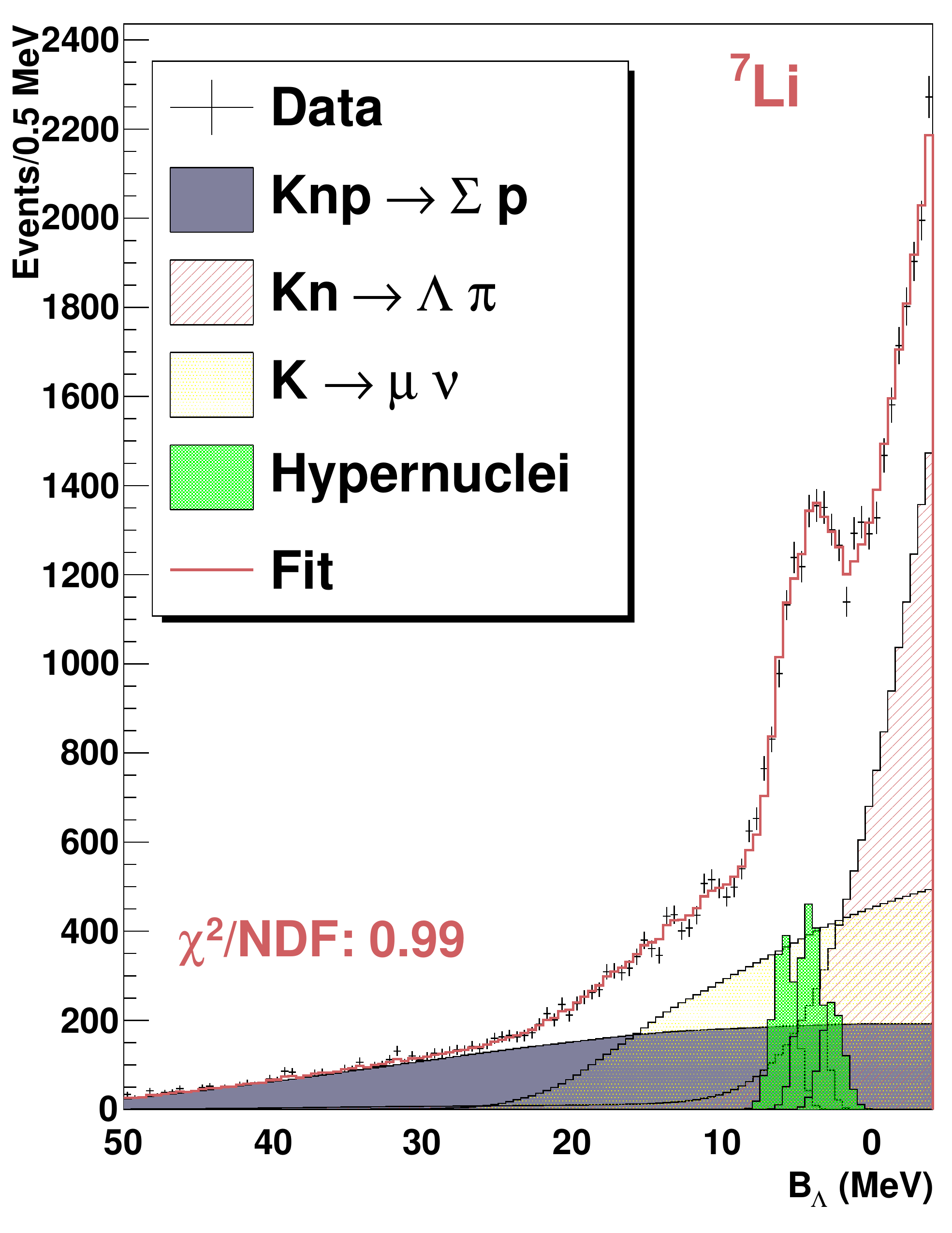}}
\caption{Binding energy distribution for the $^7$Li targets (see text for details of the fit).}
\label{f:li7}
\end{center}
\end{figure}

\begin{figure}[t]
\begin{center}
\resizebox{7.5cm}{!}{\includegraphics{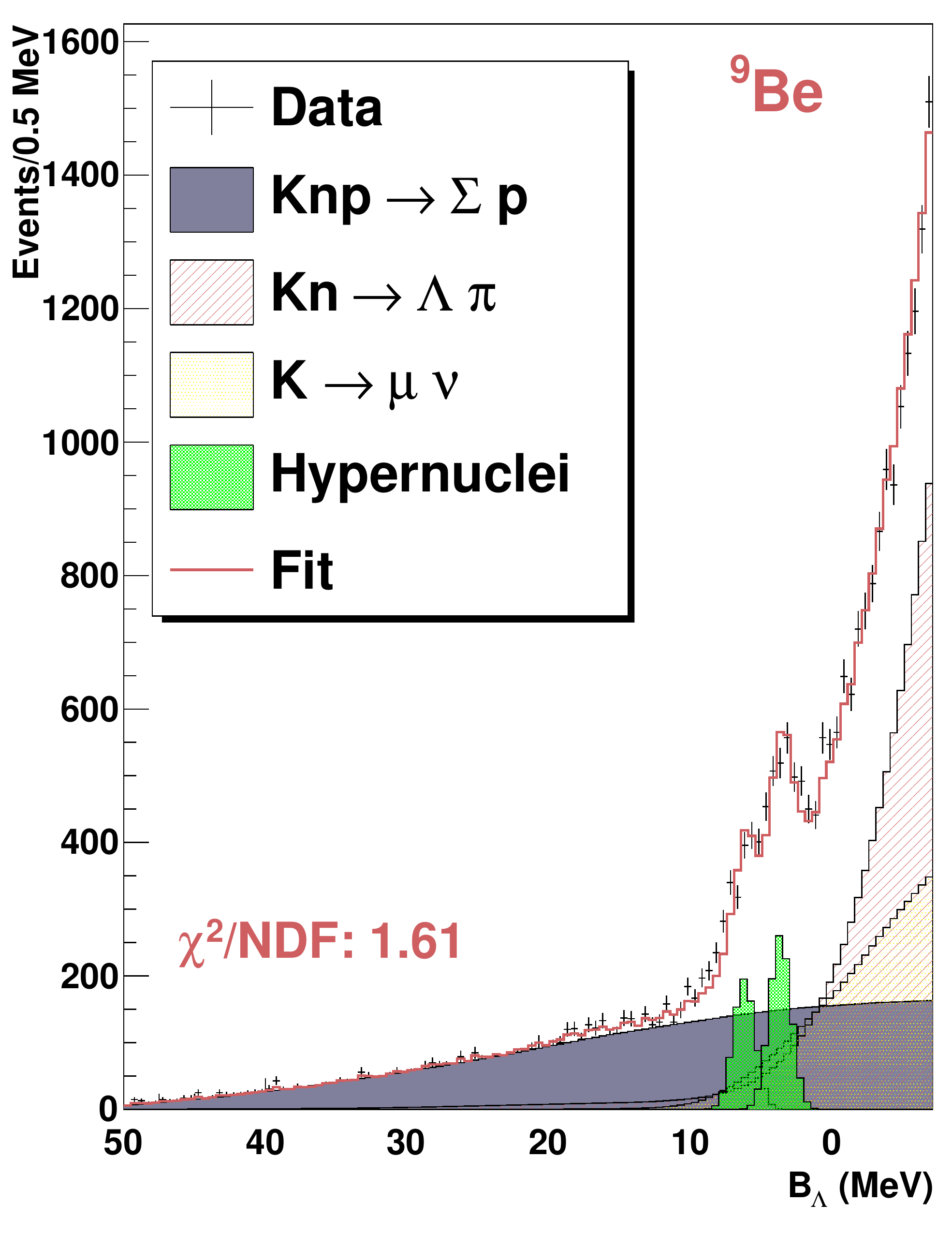}}
\caption{Binding energy distribution for the $^9$Be targets (see text for details of the fit).}
\label{f:be9}
\end{center}
\end{figure}

\begin{figure}[t]
\begin{center}
\resizebox{7.5cm}{!}{\includegraphics{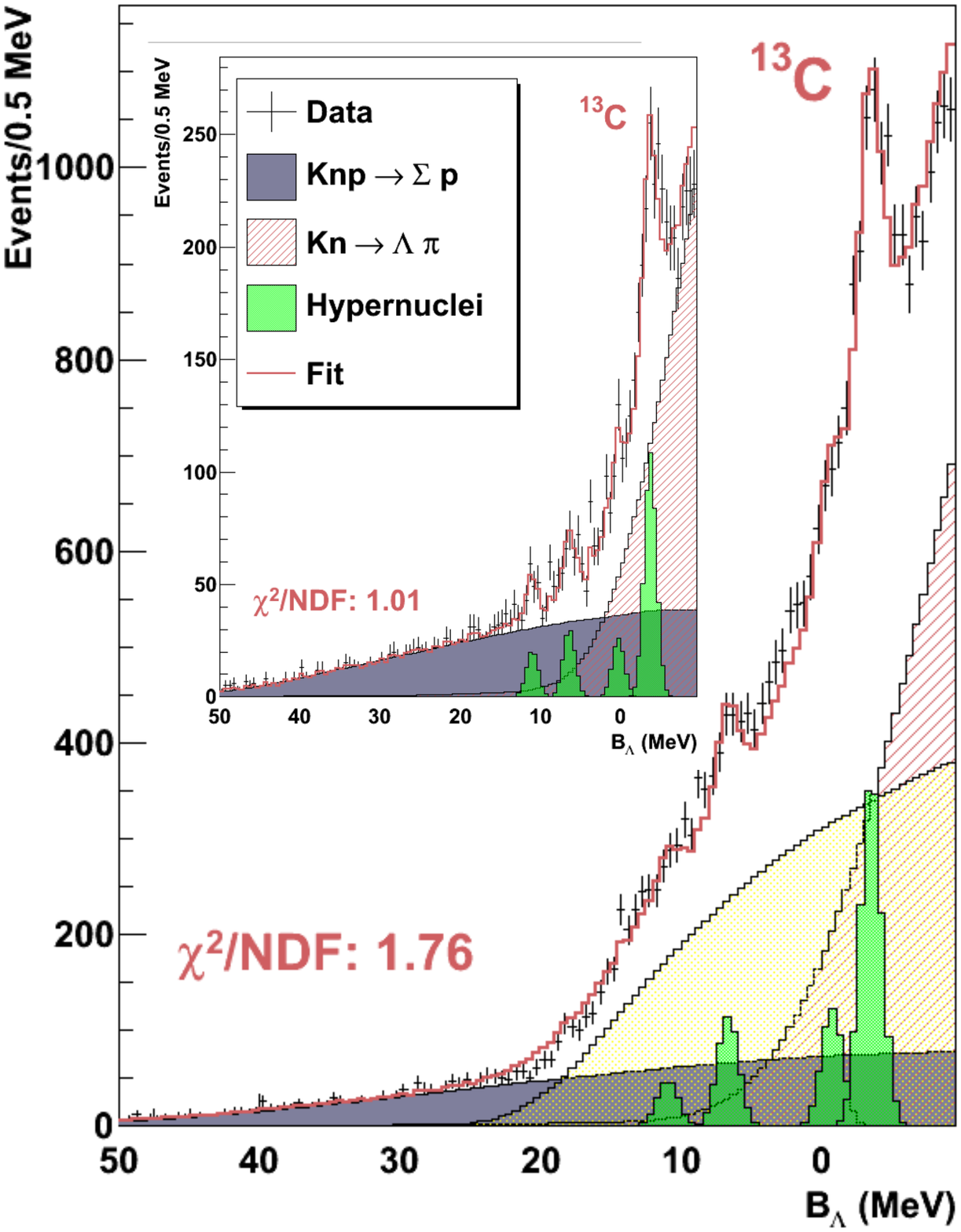}}
\caption{Binding energy distribution for the $^{13}$C targets (see text for details of the fit). In the inset  the binding energy distribution for backward tracks only is shown. In this way the background from $K^-$ in-flight decay is reduced and the ground state is more clearly visible.}
\label{f:c13}
\end{center}
\end{figure}

\begin{figure}[t]
\begin{center}
\resizebox{7.5cm}{!}{\includegraphics{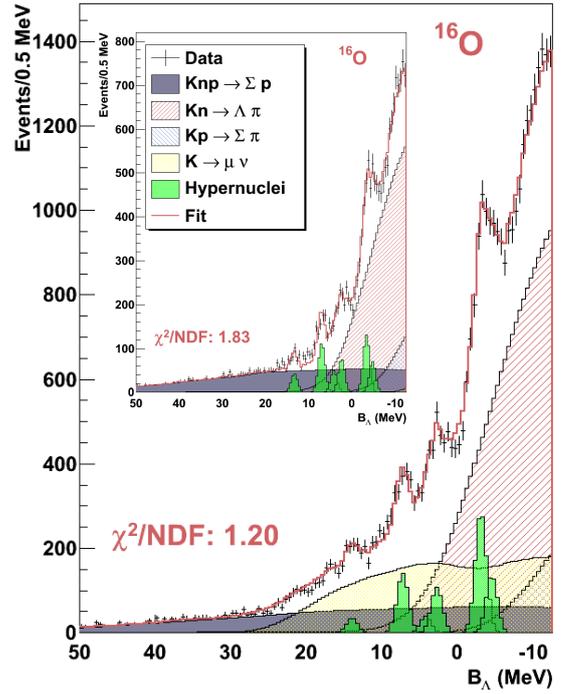}}
\caption{Binding energy distribution for the $^{16}$O targets (see text for details of the fit). In the inset  the binding energy distribution for backward tracks only is shown. In this way the background from $K^-$ in-flight decay is reduced and the ground state is more clearly visible.}
\label{f:o16}
\end{center}
\end{figure}

\section{Results}
In the following, results about the formation probability and the binding energy will be presented target by target, as a function of A: $^7$Li, $^9$Be, $^{13}$C and $^{16}$O. The number of Gaussians used to perform each of the following fits was chosen taking into account the number of clear signals in the histograms, previous experimental data, 
theoretical predictions and the $\chi^2$ of the fit.

\subsection{Formation probability and binding energy for $^7_{\Lambda}$Li}
The binding energy distribution in the bound region along with the best fit is shown in fig.~\ref{f:li7}. Even if no individual peaks are present a clear enhancement is visible in the region around 5 MeV. The best fit was obtained with three Gaussians for a total number of reconstructed events of about 4000. The position of the mean value along with the probability formation are summarized in tab.~\ref{t:li7}. The values are in agreement with a previous FINUDA publication on $^7_{\Lambda}$Li \cite{li7}, which was based on the data collected during the first data taking period with a lower statistics. These measurements can also be compared with KEK experiment E336,  that collected high statistics with the ($\pi^+, K^+$) reaction \cite{hashimoto}. For what concerns the ground state binding energy our value of $5.8 \pm 0.4$ MeV is higher than the E336 one at $5.22 \pm 0.08$ MeV. On the other hand our value agrees within 1 $\sigma$ with the accurate measurements in emulsion experiments in the sixties and seventies which reported an average value of $5.58 \pm 0.03$ \cite{davis}. For the excited states a comparison can be made with the very precise measurements performed by the Hyperball experiments \cite{ukaili7}.  Given the FINUDA experimental error on the binding energy measurements ($0.4$ MeV ) it is however difficult to affirm which of the states corresponds to the one observed in FINUDA. A reasonable hypothesis is that the state at $5.8$ MeV is the $1/2^+$ ground state. 
The second peak could be attributed to the $5/2^+$ state at $E_X = $ 2.05 MeV \cite{ukaili7}, while the third peak could represent the $T=1, \; 1/2^+$ state.
Assuming the first Gaussian contains only events from the ground state a formation probability of $(0.37 \pm 0.04 \pm 0.05)   \times 10^{-3}$ has been calculated. The sum of the rates for all the Gaussians gives a total probability of forming a bound hypernucleus per stopped $K^-$ of $(1.04 \pm 0.12 \pm 0.14)  \times 10^{-3}$. Along with the FINUDA value reported in \cite{li7} this represents the first measurement of formation probability for $^7_{\Lambda}$Li.

\begin{table}[t]
\begin{center}
\begin{tabular}{|c|c|c|c|}
\hline
 \small{\bf{$^{7}$Li}}&  \small {$B_{\Lambda}$}  &  \small {$E_X$} &  \small {Formation probability} \\
 &  \small {(MeV)}  &  \small { (MeV)} &  \small {per stopped $K^-$ ($10^{-3}$)}\\
\hline
 \small {1} &  \small{$5.8 \pm 0.4 \; $} &  \small {-} &  \small {$0.37 \pm 0.04 \pm 0.05 $}\\
 \small {2} &  \small{$4.1 \pm 0.4 \; $} &  \small {$1.7$} &  \small{$0.46 \pm 0.05 \pm 0.06 $}\\
 \small {3} &  \small{$2.6 \pm 0.4 \; $} &  \small {$3.2$} &  \small{$0.21 \pm 0.03 \pm 0.03 $}\\
\hline
\end{tabular}
\end{center}
\caption{Binding energy and formation probability for the $^7_{\Lambda}$Li states. $E_X$, as for the following tables, represents the excitation energy, that is the binding energy difference, with respect to the ground state.}
\label{t:li7}
\end{table}

\subsection{Formation probability and binding energy for $^9_{\Lambda}$Be}
The binding energy distribution in the bound region for the $^9$Be targets is shown in fig.~\ref{f:be9}. Two signals are clearly visible, one centered at $6.2$ MeV, the other at $3.7$ MeV. 
Our background, rising steeply above 0 MeV, doesn't allow the ascertainment of the existence of other exited states. The E336 experiment \cite{hashimoto} on the other hand reported the presence of eight states, the position of the first two being in agreement with our measurement. The binding energy of the ground state is somehow lower than the value of $6.71 \pm 0.04$ 
MeV measured in emulsion data \cite{davis}, but still compatible within the errors. The excitation energy of the second peak is compatible with high precision $\gamma$-spectroscopy measurements reported in \cite{tam}. A total of about 1800 reconstructed events have been found, corresponding to a formation probability of $(0.16 \pm 0.02 \pm 0.02)   \times 10^{-3} $ and of $(0.21 \pm 0.02 \pm 0.03)   \times 10^{-3}$ for the ground and the excited states respectively, for a total probability of $(0.37 \pm 0.04 \pm 0.05)   \times 10^{-3}$ of forming a  $^9_{\Lambda}$Be when stopping a $K^-$. All the results are summarized in tab.\ref{t:be9}. This is the first world measurement of formation probability for $^9_{\Lambda}$Be.
\begin{table}[t]
\begin{center}
\begin{tabular}{|c|c|c|c|}
\hline
 \small{\bf{$^{9}$Be}} &  \small {$B_{\Lambda}$}  &  \small {$E_X$} &  \small {Formation probability} \\
  &  \small {(MeV)}  &  \small {(MeV)} &  \small {per stopped $K^-$ ($10^{-3}$)}\\
\hline
 \small {1} &  \small{$6.2 \pm 0.4 \; $} &  \small {-} &  \small {$0.16 \pm 0.02 \pm 0.02$}\\
 \small {2} &  \small{$3.7 \pm 0.4 \; $} &  \small {$2.5$} &  \small{$0.21 \pm 0.02 \pm 0.03$}\\
\hline
\end{tabular}
\end{center}
\caption{Binding energy and formation probability for the $^9_{\Lambda}$Be states.}
\label{t:be9}
\end{table}

\subsection{Formation probability and binding energy for $^{13}_{\Lambda}$C}
The binding energy distribution for the $^{13}$C target is shown in fig.~\ref{f:c13}. Besides a clear peak above 0 MeV and a small peak around 7 MeV, no other hypernuclear state is visible. For this reason the high quality backward tracks sample, that has much less background from in-flight $K^-$ decays, is shown in the inset. 
Even if the statistics is lower, four peaks are visible and the ground state becomes cleaner. The overall fit has been performed with four Gaussians as suggested by theoretical 
predictions \cite{itonaga} and by the experimental distribution. A better $\chi^2$/NDF, $1.50$ instead of 1.76 of fig.~\ref{f:c13}, could be obtained with the use of 5 Gaussians, 
the additional peak positioning at $E_{X} = $ 7.6 MeV. The inclusion of this peak does not change the measured formation probabilities reported in tab.~\ref{t:c13} in any significant way. 
The peak at $B_{\Lambda} = - 3.7$ MeV is assigned to an unbound $^{13}_{\Lambda}$C state decaying to $^{12}$C $+ \Lambda$. For what concerns the third peak it is sitting very close to 0 MeV. Since the mean value is at $0.3 \pm 0.4$ MeV it will be considered in the bound region. The results of the fit in terms of Gaussians are reported in tab.~\ref{t:c13} for a total number of reconstructed events of about 1100 for the first three peaks. A comparison can be made with previous values by E336 \cite{hashimoto}. 
While similar binding energies are reported for the ground state, the first excited state and the state in the unbound region, they report the presence of two 
other peaks at $E_X =$ 9.73 and at $E_X =$ 11.75 MeV. The only precise $\gamma$ spectroscopy measurement revealed excited states at 4.88 MeV and 11 MeV \cite{koh}, compatible with our second and third peak. The ground state measured in an emulsion experiment was found at $11.69 \pm 0.12$ MeV \cite{davis}, value within 2 $\sigma$ from our measurement. For what concerns the formation probabilities the values are reported in tab.~\ref{t:c13}. For the ground state the value $(0.10 \pm 0.02 \pm 0.01)   \times 10^{-3} $ is obtained, while summing over all the three states the total formation probability is $(0.45 \pm 0.08 \pm 0.09)  \times 10^{-3}$ per stopped $K^-$. Also in this case this is the first measurement of formation probability for $^{13}_{\Lambda}$C.

\begin{table}[t]
\begin{center}
\begin{tabular}{|c|c|c|c|}
\hline
 \small{\bf{$^{13}$C}} & \small {$B_{\Lambda}$}  &  \small {$E_X$} &  \small {Formation probability} \\
 \small{} &  \small {(MeV)}  &  \small {(MeV)} &  \small {per stopped $K^-$ ($10^{-3}$)}\\
\hline
 \small {1} &  \small{$11.0 \pm 0.4 \; $} &  \small {-} &  \small {$0.10 \pm 0.02 \pm 0.01 $}\\
 \small {2} &  \small{$6.4 \pm 0.4 \; $} &  \small {$4.6$} &  \small{$0.19 \pm 0.02 \pm 0.03 $}\\
 \small {3} &  \small{$0.3 \pm 0.4 \; $} &  \small {$10.7$} &  \small{$0.16 \pm 0.02 \pm 0.02 $}\\
 \small {4} &  \small{$-3.7 \pm 0.4 \; $} &  \small {$14.7$} &  \small{$0.47 \pm 0.04 \pm 0.07 $}\\
\hline
\end{tabular}
\end{center}
\caption{Binding energy and formation probability for the $^{13}_{\Lambda}$C states.}
\label{t:c13}
\end{table}

\subsection{Formation probability and binding energy for $^{16}_{\Lambda}$O}
The binding energy distribution for the D$_2$O target is shown in fig.~\ref{f:o16}. In the inset the distribution for the backward tracks sample is also shown, where the ground state is more clearly visible. The overall distribution has been fitted to a total of six Gaussians, needed for a good $\chi^2$ to be obtained. The first two peaks are attributed to $^{16}_{\Lambda}$O states, while the others are attributed to unbound  $^{16}_{\Lambda}$O states decaying to $^{15}_{\Lambda}$N hyperfragment since the particle stability threshold in $^{16}_{\Lambda}$O is at about 7.8 MeV \cite{millener}.
As reported in tab.~\ref{t:o16}, the ground state has been found at $13.4$ MeV, while the first excited state lies $6.3$ MeV below it. For what concerns the ground state this value is in agreement with a previous measurement with stopped $K^-$ \cite{hayano} ($12.9 \pm 0.4$ MeV), while it is not compatible with the value of E336 \cite{hashimoto} ($12.42 \pm 0.05$ MeV). Another measurement has been also reported using the electroproduction ($e, e' K^+$) reaction on $^{16}$O leading to the formation of $^{16}_{\Lambda}$N ($13.76 \pm 0.16$ MeV) \cite{cusanno}, in agreement with our result. The excitation energy of $6.3$ MeV is in agreement with the high precision $\gamma$ spectroscopy  performed in Hyperball experiments \cite{uka} that found a doublet at 6.562 and 6.786 MeV above the ground state.The total number of reconstructed events in the first two peaks amounts to about 750. The formation probability for the ground state and the first excited state have been measured to be respectively $(0.10 \pm 0.02 \pm 0.01)   \times 10^{-3} $ and $(0.26 \pm 0.04 \pm 0.04)   \times 10^{-3} $, for a total hypernucleus formation of $(0.36 \pm 0.06 \pm 0.05)  \times 10^{-3}$ per stopped $K^-$.  These values are compatible with those measured previously \cite{hayano}. 

\begin{table}[t]
\begin{center}
\begin{tabular}{|c|c|c|c|}
\hline
 \small{\bf{$^{16}$O}} & \small {$B_{\Lambda}$}  &  \small {$E_X$} &  \small {Formation probability } \\
 \small{} &  \small {(MeV)}  &  \small {(MeV)} &  \small {per stopped $K^-$ ($10^{-3}$)}\\
\hline
 \small {1} &  \small{$13.4 \pm 0.4 \; $} &  \small {-}         &  \small {$0.10 \pm 0.02 \pm 0.01 $}\\
 \small {2} &  \small{$7.1 \pm 0.4 \; $} &  \small {$6.3$}   &  \small{$0.26 \pm 0.04 \pm 0.04 $}\\
 \small {3} &  \small{$4.3 \pm 0.4 \; $} &  \small {$9.1$}   &  \small{$0.13 \pm 0.03 \pm 0.02 $}\\
 \small {4} &  \small{$2.4 \pm 0.4 \; $} &  \small {$11.0$}  &  \small{$0.15 \pm 0.03 \pm 0.02 $}\\
 \small {5} &  \small{$-3.3 \pm 0.4 \; $} &  \small {$16.7$} &  \small{$0.55 \pm 0.07 \pm 0.08 $}\\
 \small {6} &  \small{$-4.7 \pm 0.4 \; $} &  \small {$18.1$} &  \small{$0.28 \pm 0.06 \pm 0.04 $}\\
\hline
\end{tabular}
\end{center}
\caption{Binding energy and formation probability for the $^{16}_{\Lambda}$O states.}
\label{t:o16}
\end{table}

\section{Discussion of the results and conclusions}
As discussed in the previous section, ref. \cite{hayano} reported measurements of formation probability with stopped $K^-$ for three type of target elements, $^4$He, $^{12}$C and $^{16}$O. Probabilities for the ground state 
formation have been found to be $(17.9 \pm 1.5)   \times 10^{-3}$ for $^4_{\Lambda}$He, $(0.98 \pm 0.12)   \times 10^{-3}$ for $^{12}_{\Lambda}$C and $(0.13 \pm 0.04)   \times 10^{-3}$ for $^{16}_{\Lambda}$O. In 2005 FINUDA \cite{C12} reported a probability of $(1.01 \pm 0.11 \pm 0.10)   \times 10^{-3}$ for the ground state of $^{12}_{\Lambda}$C. Ref. \cite{ahmed} also measured the hypernuclei formation probability in the ($K^-_{stop},\; \pi^o$) reaction using a $^{12}$C target and reported a value of $(0.28 \pm 0.08)   \times 10^{-3}$ for the ground state, calculated on a limited sample of 13.7 $\pm$ 4 events. Based on isospin conservation, this value must be multiplied by two to be compared with the previous ones measured in ($K^-_{stop},\; \pi^-$) production experiments.

From these set of data, it appears that the formation probability is a decreasing function of the atomic mass number A but the overall frame is not coherent, especially due to the difference between the ground state formation probabilities measured by \cite{hayano,C12} and by \cite{ahmed}. The new measurements reported here give a more complete picture of the situation, since they report for the first time also probabilities for $^7_{\Lambda}$Li,  $^9_{\Lambda}$Be and $^{13}_{\Lambda}$C. These values can be compared directly to the one reported previously by FINUDA \cite{C12} since they are measured in the same experiment and using the same experimental and reconstruction techniques. The relative behavior is thus free from possible systematic errors in comparing values from different experiments.

\begin{figure}[t]
\begin{center}
\resizebox{7.5cm}{!}{\includegraphics{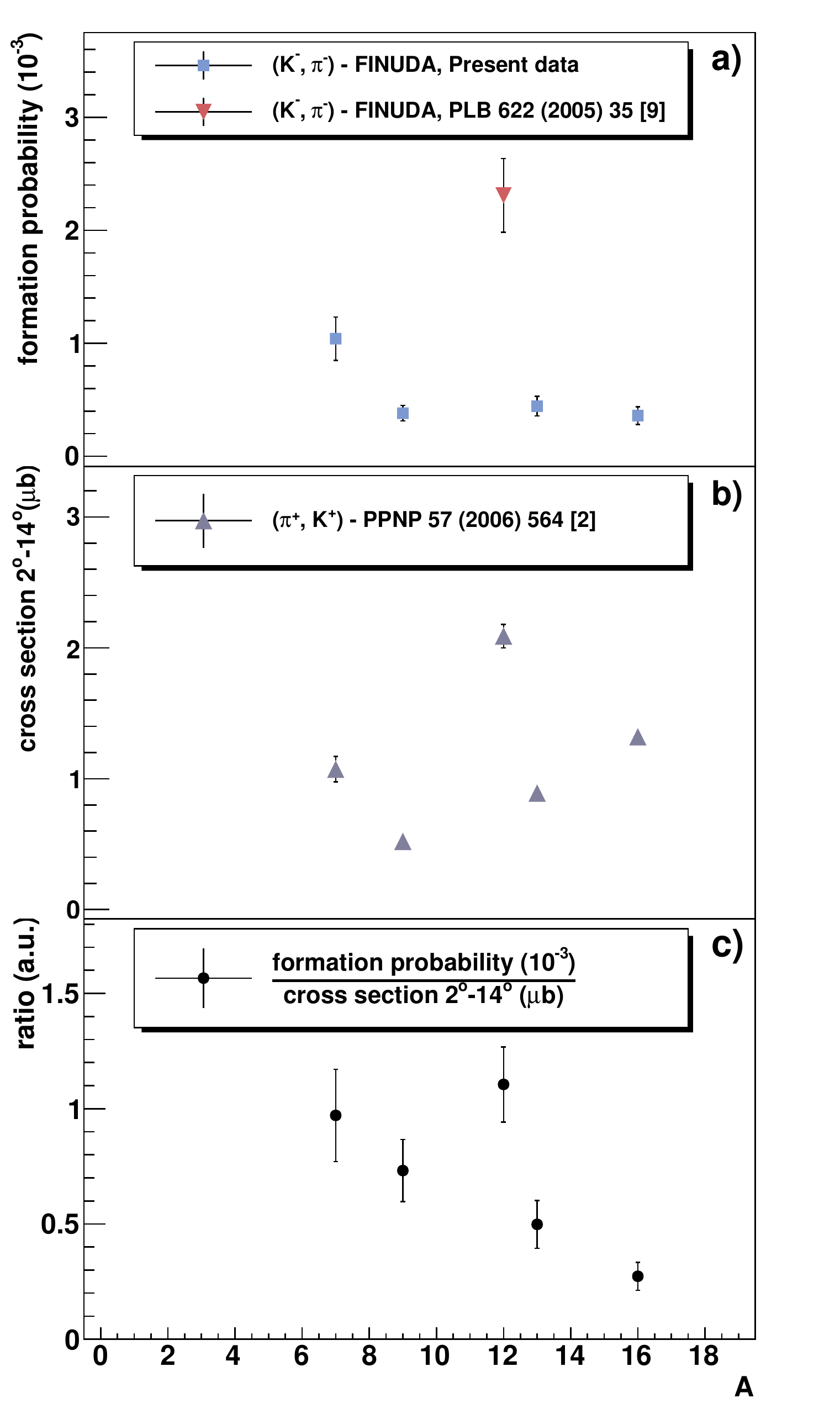}}
\caption{Formation probabilities from FINUDA (a) and cross section from E336 \cite{hashimoto} (b) for bound states, see text for details. In c) the ratio between the two is shown.}
\label{f:vsA}
\end{center}
\end{figure}

It is not easy to draw a simple A dependence from the above data on the $K^-$ capture probabilities for the examined p-shell hypernuclei. As a matter of fact, if we considered only the ground states of them, there are strong differences in the description of the nuclear configurations, and in some cases a not completely clean separation, from an experimental point of view, of the ground state from low-lying excited states (e.g. $^7_{\Lambda}$Li or $^{12}_{\Lambda}$C, in which the doublet of states, one of which is the ground state, has a spacing 
of some hundreds of keV). On the other hand, if we considered all states with $B_{\Lambda}>0$, we know that in some cases such as for $^{16}_{\Lambda}$O, peaks at  $B_{\Lambda}$ = $4.3$ and $2.4$ MeV are interpreted as due to the formation of the $^{15}_{\Lambda}$N+p system.
For these reasons, in order to consider only well defined hypernuclides, we selected only hypernuclear states with energy below the threshold for the decay by proton emission. Explicitly it means states 1, 2 and 3 for $^7_{\Lambda}$Li, 1 and 2 for $^{9}_{\Lambda}$Be, 1, 2, 3 and 4 for $^{12}_{\Lambda}$C (fit 1, Tab. I of \cite{C12}), 1, 2 and 3 for $^{13}_{\Lambda}$C and 1 and 2 for $^{16}_{\Lambda}$O (see tabs. \ref{t:li7} to \ref{t:o16}). Fig.~\ref{f:vsA}a) shows a plot of the capture rates chosen following the above criterion as a function of A. A smoothly decreasing behaviour appears, with the exception of a strong enhancement corresponding to the formation of $^{12}_{\Lambda}$C bound states.
The only other experiment which studied the same targets with a comparable energy resolution, a comparable momentum transfer (about 300 MeV/c), observing a similar pattern of excited states, was the aforementioned E336 experiment with the ($\pi^+,K^+$) reaction at 1.05 GeV/c \cite{hashimoto}. Fig.~\ref{f:vsA}b) shows the differential cross section integrated in the forward direction (2$\mathrm{^o}$-14$\mathrm{^o}$) for the same peaks, and fig.~\ref{f:vsA}c) the ratio between the two values. 
This ratio ranges by a factor close to five in the same p-shell hypernuclei, from a large value for $^{7}_{\Lambda}$Li and $^{12}_{\Lambda}$C to a small value for $^{16}_{\Lambda}$O, showing a distinct A dependence for the two reactions, $K^-$ capture at rest and in-flight ($\pi^+,K^+$).\\

In conclusion we have reported the first measurements of the formation probabilities of different hypernuclear states following the capture at rest of $K^-$ from $^{7}$Li, $^{9}$Be and$^{13}$C targets, as well as a new measurement for the $^{16}$O target. Together with previous measurements on $^{12}$C target, this bank of data allows for a meaningful study of the formation of p-shell hypernuclei from the two-body capture of $K^-$ at rest. The possibility of disentangling the effects due to atomic wave-function of the captured $K^-$ from those due to the pion optical nuclear potential and from those due to the specific hypernuclear states can be achieved by a combined analysis of several hypernuclides, 
as shown in the following Letter \cite{cieply}.

\section{Acknowledgements}
The authors would like to thank Avraham Gal for the fruitful discussions about the interpretation of the data presented in this Letter.

\end{document}